# A two-stage full-band speech enhancement model with effective spectral compression mapping

Zhongshu HOU[1,2,3]; Qinwen HU[1,2,3]; Kai CHEN[1,2,3]; Jing LU[1,2,3]

[1] Key Laboratory of Modern Acoustics, Nanjing University, Nanjing 210093, China

[2] NJU-Horizon Intelligent Audio Lab, Horizon Robotics, Beijing 100094, China

[3] Nanjing Institute of Advanced Artificial Intelligence, Nanjing 210014, China

**ABSTRACT**

The direct expansion of deep neural network (DNN) based wide-band speech enhancement (SE) to full-band processing faces the challenge of low frequency resolution in low frequency range, which would highly likely lead to deteriorated performance of the model. In this paper, we propose a learnable spectral compression mapping (SCM) to effectively compress the high frequency components so that they can be processed in a more efficient manner. By doing so, the model can pay more attention to low and middle frequency range, where most of the speech power is concentrated. Instead of suppressing noise in a single network structure, we first estimate a spectral magnitude mask, converting the speech to a high signal-to-ratio (SNR) state, and then utilize a subsequent model to further optimize the real and imaginary mask of the pre-enhanced signal. We conduct comprehensive experiments to validate the efficacy of the proposed method.

## 1. INTRODUCTION

Speech enhancement (SE) plays an important role in front-end processing for many applications such as speech communication, automatic speech recognition (ASR) and digital hearing aids, aiming at improving the overall perceptual quality and intelligibility of speech signals distorted by background disturbances. In the last decade, data-driven deep neural network (DNN) based SE methods have achieved significantly better performance over traditional rule-based signal processing SE methods especially in adverse environments with non-stationary interference (1). However, most of the DNN-based methods are designed for signals with 16 kHz sampling rate, and the quality of speech signal at this sampling rate is still inferior to the full band speech with 48 kHz sampling rate.

Most DNN based SE methods operate in the time-frequency (T-F) domain to estimate a mask between clean and noisy spectrum (2, 3) or directly predict the real and imaginary parts of the target clean complex spectrum from the noisy speech (4, 5). There are also methods directly estimating the raw-waveform of the clean signals in the time domain (6). Recently, convolutional recurrent network (CRN) was proposed (7) and has been proven an efficient network structure for SE in T-F domain with low computational complexity. It utilizes the convolution neural network (CNN) to capture the local features in spectrogram and the recurrent neural network (RNN) to exploit the temporal correlation between consecutive frames. The dual-path recurrent neural network (DPRNN) (8), originally designed to overcome the deficiency of conventional RNN in modeling long sequence in temporal dimension, can also be applied in the frequency dimension with the inherent advantage of making full use of the harmonic spectral structure of speech. We have combined the benefits of CRN and DPRNN and designed a model called dual-path convolutional recurrent network (DPCRN) (9).

Self-attention mechanism, first introduced in (10), is broadly utilized in sequence-to-sequence tasks. It can more efficiently model long-term dependencies than RNNs (11) and temporal convolutional networks (TCNs) (12, 13) with more efficient parallelization capability . Recent studies have shown that the attention mechanism also performs well in SE tasks (12, 14).

* This work was supported by National Science Foundation of China with Grant No. 11874219.

Most of the SE models are basically employed in wide band (16 kHz) scenarios, while their performance has not been validated in full band (48 kHz) processing. The direct expansion of these models to full band is not reasonable. The threefold computational complexity will hinder its real-time implementation. More importantly, uniform processing in the frequency domain will comparatively weaken the network's modeling capacity in low and middle frequency range where the majority of speech power concentrates. The perceptual band focusing on spectral envelope of speech (15) is a possible solution, but the rough resolution of spectrum features may drop important information and degrade the recovered speech quality.

Inspired by the chain optimization framework (5), we propose a two-stage full band SE model called MHA-DPCRN in this paper. In the first stage, a multi-head attention network (MHAN) is trained to estimate the amplitude mask between the noisy spectrogram and the clean spectrogram, aiming at converting the noisy speech to a relatively high SNR state. In the second stage, a DPCRN structure is jointly trained with MHAN to further optimize the real and imaginary parts of the complex spectrum. In both stages, a learnable spectral compression mapping (SCM) is designed to keep vital speech components intact while smoothly compressing the high frequency bands. Accordingly, a learnable inverse spectral compression mapping (iSCM) is used to reconstruct the full-band spectrum and compensate the information loss during the compressing procedure. We conduct comprehensive experiments on different open datasets to demonstrate the advantage of the SCM/iSCM structure and the efficacy of the proposed model.

## 2. MHA-DPCRN

### 2.1 Problem Formulation

Let $S(n,k)$ and $X(n,k)$ denote clean and noisy speech in the T-F domain at time frame index $n$ and frequency bin index $k$. For simplicity, indexes $n$ and $k$ are omitted hereafter. To recover the clean speech from the degraded signal, an MHAN, denoted as $\mathcal{F}_{MHA}$, is first applied to estimate the spectral magnitude mask (SMM) (16) $\mathcal{M}$ and the operation can be expressed as

$$\mathcal{M} = \mathcal{F}_{MHA}(|X|) \qquad (1)$$

$$\tilde{S}_{MHA} = \mathcal{M} \odot X \qquad (2)$$

where $\odot$ denotes element-wise multiplication and $\tilde{S}_{MHA}$ is the enhanced spectrogram by MHAN. To further suppress the residual noise and retrieve the phase magnitude details, a successive network on DPCRN is employed to directly estimate the real and imaginary parts of the complex spectrogram. Then the enhancement process can be represented as:

$$\tilde{S}_{DPCRN} = \mathcal{F}_{DPCRN}(\tilde{S}_{MHA}) \qquad (3)$$

where $\mathcal{F}_{DPCRN}$ and $\tilde{S}_{DPCRN}$ denote the operation of DPCRN and the final enhanced speech spectrogram respectively.

### 2.2 Model Architecture

As shown in Figure 1(a), the proposed two-stage network consists of two principal parts, namely MHAN and DPCRN, and the network details are as follows.

#### 2.2.1 Learnable Spectral Compression Mapping

Directly expanding the bandwidth of a wide-band network to full-band with the same spectral resolution will allocate two thirds of the computational resources to the less informative high band (8 kHz - 24 kHz), resulting in significant increase of both the computational burden and the learning difficulty. Therefore, it is necessary to introduce a spectral information mapping strategy to effectively compress the high band. Mel-scale filter banks are widely used to extract Mel-frequency cepstral coefficients in speaker recognition (17) due to their analogy to the behavior of human auditory system. Similar to the Mel-scale, our converted scale is defined with a logarithmic function. To further retain the information in the critical low band, we warp the spectrum by keeping frequencies below 5 kHz intact while only transforming frequencies above 5 kHz logarithmically, with the mapping curve defined as

$$f_c = \begin{cases} f & ,0 \leq f \leq 5kHz \\ 2500\left[\ln(\frac{f-2500}{2500}) + 2\right] & ,5k < f \leq 24kHz \end{cases}. \tag{4}$$

Based on the mapping in equation (4), the compression operation is designed. Suppose the $F$-dimensional spectrum is compressed into an $F_c$-dimensional spectral representation, with the first $K$ dimensions denoting the uncompressed low frequencies. The compression can be realized using a matrix multiplication between a compression matrix of size $F_c \times F$ and the original spectrum. The $K \times F$ low band mapping portion of the compression matrix keeps low frequency bands unchanged with a $K \times K$ identity submatrix, while the $(F_c - K) \times F$ high-band part initialized with a series of triangular filter banks compresses higher bands logarithmically (18). Note that the mapping of equation (4) roots from the human auditory system and cannot effectively match the sparse distribution of speech spectrum in high frequency range. Therefore, the direct implementation of this compression pattern would lead to considerable residual noise.

To more effectively exploit the sparse distribution of speech in high frequency range, we propose to set the compression matrix partially learnable. In network implementation, a dense layer with no bias is initialized with weights of the compression matrix. The low band mapping of layer is fixed and the high-band part is learned by the network. Correspondingly, the inverse spectral compression mapping (iSCM) is also through a learnable dense layer, but it is randomly initialized. It is observed that the inversion pattern is closely related to the compression pattern.

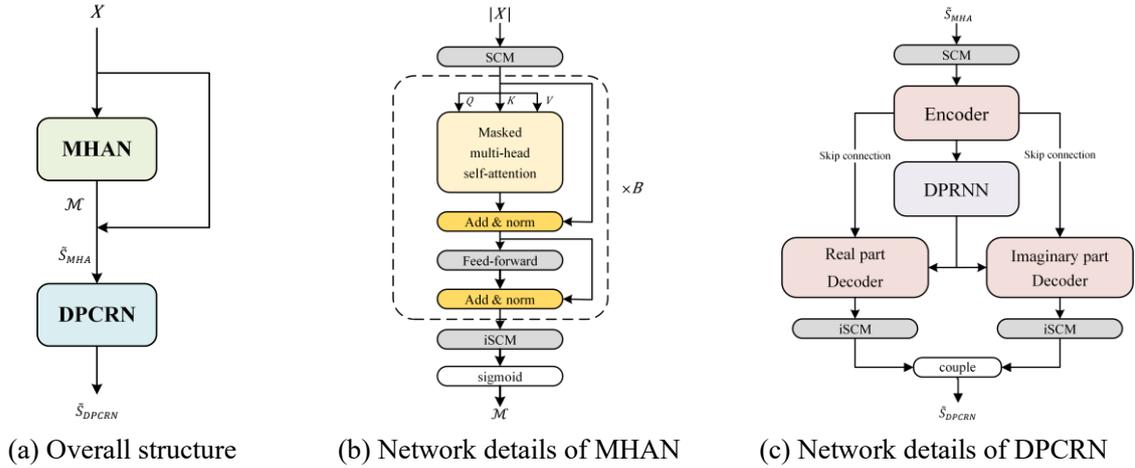

(a) Overall structure     (b) Network details of MHAN     (c) Network details of DPCRN

Figure 1 – Overall diagram of the proposed framework

### 2.2.2 Multi-Head Attention Network

As shown in Figure 1(b), MHAN resembles the multi-head attention network structure used in (12). It first employs the aforementioned SCM to compress frequency bands in each frame to an $F_c$-dimension vector, along with a frame-wise normalization and a ReLU activation layer. $B$ repetitive MHA blocks follow afterwards and no positional encoding is applied. An MHA block consists of a masked MHA module and a two-layer feed-forward module, with a residual connection (19) and a frame-wise layer normalization applied after each module. The detailed description on MHA can be found in (12, 13). Finally, the processed information goes through an iSCM and is mapped into an interval between 0 and 1 by a sigmoid function, generating the SMM. The product of the estimated SMM and input STFT spectrogram is obtained in an element-wise manner, and then fed to the DPCRN.

### 2.2.2 Dual-Path Convolutional Recurrent Network

DPCRN consists of an encoder, a dual-path RNN (DPRNN) module and two decoders, as shown in Figure 1(c). The real and imaginary parts of the noisy spectrogram are mapped by an SCM before sending to the encoder. The encoder utilizes the 2-dimensional convolutional (Conv-2D) layers to extract local patterns from noisy spectrogram. The decoders use the transposed 2-dimensional convolutional layers (TransConv-2D) symmetric with that in the encoder to refactor low-resolution features to the original shape. There are skip connections between the encoder and the decoders for information interaction. Every convolutional layer is followed by a batch normalization layer and a PReLU activation function. In the DPRNN module, inter-chunk processing uses long short-term memory (LSTM) unit to model the temporal dependence. As for the intra-chunk processing, bidirectional LSTM (BiLSTM) is employed to model spectral patterns in a single frame. The LSTM and BiLSTM are followed by a dense layer and an instance normalization (IN). A residual connection is then applied between the input of RNN and the output of IN to further mitigate the gradient vanishing problem.

**2.2.3 Training Targets and Loss Function**

Power compress loss functions (20) are utilized to better process the information in low power T-F points:

$$L_{RI}(\tilde{S},S) = \left|\left|S^{\mathcal{C}}_{real} - \tilde{S}^{\mathcal{C}}_{real}\right|\right|^2_F + \left|\left|S^{\mathcal{C}}_{imag} - \tilde{S}^{\mathcal{C}}_{imag}\right|\right|^2_F, L_{Mag}(\tilde{S},S) = \left|\left||S|^{\gamma} - |\tilde{S}|^{\gamma}\right|\right|^2_F \quad (5)$$

$$S^{\mathcal{C}}_{real} = |S|^{\gamma}\cos\theta_S, \; S^{\mathcal{C}}_{imag} = |S|^{\gamma}\sin\theta_S, \quad (6)$$

where $\theta_S$ denotes the phase angle of complex spectrogram, $\gamma$ refers to the compression parameter, superscript $\mathcal{C}$ denotes the power compressed pattern, and $||\cdot||_F$ refers to the Frobenius norm of the matrix. MHAN is pre-trained based on the loss function below to estimate an SMM:

$$\mathcal{L}_1 = \mathcal{L}_{Mag}(\tilde{S}_{MHA}, S). \quad (7)$$

In the joint training procedure, MHAN is initialized with checkpoint pre-trained and the parameters of MHAN and DPCRN are optimized simultaneously with the loss function

$$\mathcal{L}_2 = \mathcal{L}_{Mag}(\tilde{S}_{MHA}, S) + \mathcal{L}_{Mag}(\tilde{S}_{DPCRN}, S) + \mathcal{L}_{RI}(\tilde{S}_{DPCRN}, S). \quad (8)$$

## 3. Experiments

### 3.1 Datasets

To demonstrate the efficiency of the SCM/iSCM structure, we first train our model on simulated datasets where clean speech clips are mainly generated from VCTK (21) and SIWIS (22), and noise recordings are from DEMAND (23) and QUT-NOISE (24). 16000 clips of clean speech (around 45 h) are generated with 8% for validation. Audios are convolved with room impulse responses randomly selected from openSLR26 and openSLR28 (25) to simulate reverberant environments. The SNR of noisy speech ranges from 15 to -5dB. In the testing stage, we select clean speech from DAPS (26) and noise from Saki (27, 28), to create simulated noisy speech. The SNR range of the test noisy speech is the same as the training set. All the audio clips used is sampled at 48 kHz.

For evaluation on real acoustic environment, we train our model with dataset provided by ICASSP 2022 DNS4 challenge (29). The training configures are the same as simulated datasets. To compare our model with previous state-of-the-art (SOTA) full-band and super-wideband SE methods, we also train and test our model on the classic open VCTK-DEMAND dataset (30).

### 3.2 Parameter Setup and Training Strategy

The window length and hop size are 25ms and 12.5ms respectively. The FFT length is 1200 and the hanning widow is used. The SCMs map the 601-dimension spectrum to a 256-dimension feature. For MAHN part, we set parameters $B = 5$ and the number of heads is 8. In DPCRN, the output channel of the Conv-2D in the encoder is [16,32,48,64,80]. The kernel size and the stride are respectively set to [(5,2),(3,2),(3,2),(3,2),(2,1)] and [(2,1),(1,1),(1,1),(1,1),(1,1)] in frequency and time dimension. We use DPRNN module with 1 intra chunk and 1 inter chunk, and the hidden size is set to 127. There are two ways to train the SCMs on simulated datasets, with one setting the whole parameters learnable while another setting the high-band part learnable, labeled as High Learn.

Warmup strategy (12) is critical in training MHAN, where the learning rate $\alpha$ is updated with the rule: $\alpha = \frac{1}{\sqrt{C}} \times \min\left(\frac{1}{\sqrt{\varphi}}, \frac{\varphi}{\sqrt{\Psi^3}}\right)$, with $C = 128$, warmup steps $\Psi = 10000$ and $\varphi$ denoting the training step. We train the model by the warmup-based Adam optimizer with $\beta_1 = 0.9$, $\beta_2 = 0.98$, $\epsilon = 10^{-9}$. The compression parameter $\gamma$ is $\frac{1}{3}$. The total parameter of the whole framework is 5.00 million.

### 3.3 Baselines and Evaluation Metrics

On the simulated test set, we compare our model with the models in DNS challenge, including NSNet2 (31), DPCRN (9). To prove the validity of the proposed SCM, we intentionally add the SCM/iSCM structure to the conventional DPCRN, marked as SCM-DPCRN. On VCTK-DEMAND test set, we compare our model with previous SOTA systems, including RNNoise (32), PercepNet (15),

DCCRN (33), DCCRN+ (34), DeepFilterNet (35), S-DCCRN (36) and DMF-Net (37). On the DNS4 blind test set, we use the NSNet2 as baseline.

We use DNS-MOS P.835 (38) to evaluate results on simulated test set. It provides simulated subjective evaluation scores namely speech quality (SIG), background noise quality (BAK), and overall audio quality (OVRL) based on a deep learning model. We also use a common objective metric called signal to distortion ratio (SDR) (39) for full-band evaluation. On the VCTK-DEMAND test set, we use the perceptual evaluation of speech quality (PESQ) (40), short-time objective intelligibility (STOI) (41) to evaluate the SE performance compared with previous SOTA methods. On the DNS4 blind test set, ITU-T P.835 (42) based subjective MOS is employed. It should be noted that the audios evaluated with PESQ, STOI and DNS-MOS metrics are all down-sampled to 16 kHz.

### 3.4 Result and Analysis

The performance on simulated test set is presented in Table 1.

Table 1 – Results on simulated test set

| Models | High Learn | BAK | SIG | OVLR | SDR(dB) |
| --- | --- | --- | --- | --- | --- |
| Noisy | | 3.175 | **4.302** | 3.289 | 1.32 |
| NSNet2 | | 4.312 | 3.952 | 3.600 | 11.37 |
| DPCRN | | 4.400 | 4.104 | 3.692 | 13.70 |
| SCM-DPCRN-1 | No | 4.414 | 4.126 | 3.735 | 14.46 |
| SCM-DPCRN-2 | Yes | 4.466 | 4.184 | 3.811 | 14.73 |
| MHA-DPCRN-1 | No | 4.502 | 4.232 | 3.874 | 15.44 |
| MHA-DPCRN-2 | Yes | **4.507** | 4.237 | **3.884** | **15.74** |

It can be seen that DPCRN with SCM structure performs better than the original DPCRN in terms of all metrics. It can be attributed to the fact that SCM makes the harmonics feature in low and middle frequency range more discriminative, as shown in Figure 2, which facilitates more effective attenuation of noise. Comparing two SCM-DPCRN networks, we can see that the one trained with High Learn strategy achieves better results. We present the learnt SCM parameters of both in Figure 3, from which we can see that the learnt SCM parameters of model SCM-DPCRN-1 will map the high-frequency information to the low-frequency range, resulting in additional distortion to major speech components. While SCM-DPCRN-2 keeps the spectrum below 5 kHz completely intact and only higher frequency bands are adjusted. The proposed two-stage MHA-DPCRN models show significant improvements over baselines. It can be seen from Figure 4 that a prior estimated SMM can effectively reduce major noise components and a subsequent network refines the spectrogram details.

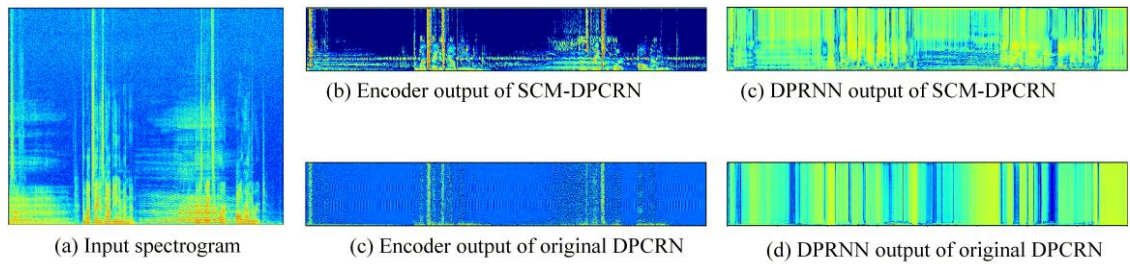

Figure 2 – Outputs comparison of DPCRN and SCM-DPCRN

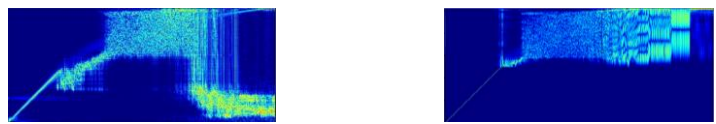

(a) Learnt SCM of SCM-DPCRN-1    (b) Learnt SCM of SCM-DPCRN-2

Figure 3 –Comparison of the learnt SCMs

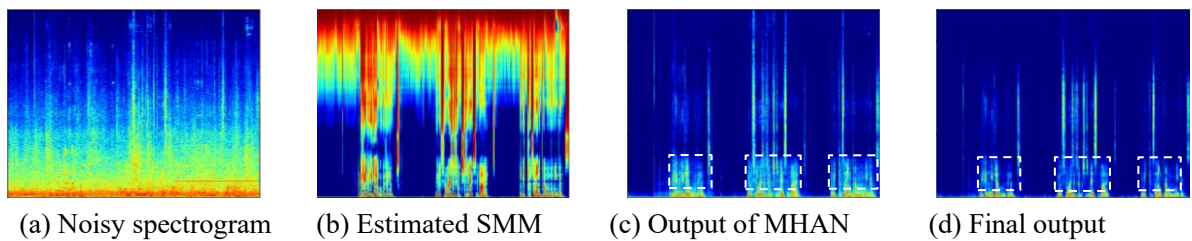

| (a) Noisy spectrogram | (b) Estimated SMM | (c) Output of MHAN | (d) Final output |

Figure 4 – Illustrating spectrograms of the enhancement process

Results on VCTK- DEMAND test set is shown in Table 2, and the proposed method achieves higher scores in terms of both metrics compared with other SOTA full-band models. On average, no less than 0.05 increase in PESQ is achieved. Compared with RNNosie and PercepNet, MHA-DPCRN yields significant improvement in PESQ and STOI, indicating that the proposed SCM/iSCM structure better improves speech quality and intelligibility by keeping the resolution of low frequency bands unchanged.

Table 2 – Results on VCTK-DEMAND test set

| Models | year | Para.(M) | PESQ | STOI(%) |
|---|---|---|---|---|
| Noisy | - |  | 1.97 | 92.1 |
| RNNoise | 2020 | 0.06 | 2.34 | 92.2 |
| PercepNet | 2020 | 8.00 | 2.73 | - |
| DCCRN | 2020 | 3.70 | 2.54 | 93.8 |
| DCCRN+ | 2021 | 3.30 | 2.84 | - |
| DeepFilterNet | 2021 | 1.80 | 2.81 | - |
| S-DCCRN | 2022 | 2.34 | 2.84 | 94.0 |
| DMF-Net | 2022 | 7.84 | 2.97 | 94.4 |
| MHA-DPCRN(**Pro.**) | 2022 | 5.00 | **3.02** | **94.4** |

The P.835 based subjective MOS on the DNS challenge blind test set is shown in Table 3, including subjective speech (SIG), background noise (BAK), overall MOS (OVRL) scores. It shows that our model still achieves remarkable performance in real acoustic scenarios.

Table 3 – Results on DNS4 challenge blind test set

| Models | BAK | SIG | OVLR |
|---|---|---|---|
| Noisy | 2.15 | **4.29** | 2.63 |
| NSNet2 | 3.93 | 3.62 | 3.26 |
| MHA-DPCRN(**Pro.**) | **4.42** | 3.97 | **3.72** |

## 4. Conclusions

We design a compression structure in spectrum called SCM for full-band SE task, which has the benefit of efficiently compressing the high frequency components while guaranteeing the effective modeling capacity in low and middle frequency range where the majority of speech power concentrates. We propose a two-stage full band speech enhancement model with SCM in the T-F domain, named MHA-DPCRN. It suppresses major disturbances with an SMM estimated by MHAN and further clear residual noise by DPCRN. With 5.0M parameters, our model achieves competitive results on various datasets.